# Error Growth Dynamic and Predictability of Tropical Cyclone in Machine Learning Weather Prediction Model

Jingchen Pu[1], Mu Mu*[1,2,3], Jie Feng[1,2,4], Hao Li[5]

[1] Department of Atmospheric and Oceanic Sciences and Institute of Atmospheric Sciences, Fudan University, Shanghai, China

[2] Shanghai Key Laboratory of Ocean-land-atmosphere Boundary Dynamics and Climate Change, Shanghai, China

[3] Shanghai Frontiers Science Center of Atmosphere–Ocean Interaction, Shanghai, China

[4] Shanghai Academy of Artificial Intelligence for Science, Shanghai, China

[5] Artifcial Intelligence Innovation and Incubation Institute, Fudan University, Shanghai, China

Corresponding author: Mu Mu (mumu@fudan.edu.cn, https://orcid.org/0000-0001-7174-817X)

22 Mar, 2026

**Abstract**

Predictability analysis, which focuses on perturbation growth dynamic, is a key problem in both weather and climate prediction. Among all perturbations, the conditional nonlinear optimal perturbation (CNOP) leads to maximum uncertainties in forecasts, which is fundamentally important for theoretical studies and applications. Traditionally, CNOPs are solved through iterative optimization of numerical weather prediction (NWP) systems. Their large computational demands pose significant challenges to long-term predictability analysis. In our study, using a fast and accurate Artificial intelligence (AI) model, i.e. FuXi, a low-cost optimization framework for solving 5-day tropical cyclone (TC) CNOP is developed. For the first time, CNOPs that achieve the optimal (i.e., fastest) nonlinear development of long-term TC forecast errors are solved, with their optimality and physical explainability verified.

Results demonstrate that perturbations with specific spatial structures undergo significant development. In both AI and NWP models, AI-based CNOPs exhibit rapid and physically consistent error growth across diverse TC cases, faster compared to random and lagged forecast perturbations. Furthermore, sensitivity analysis reveals that far-environment systems and processes are more crucial for long-term TC forecasts. Structural analyses of the CNOP emphasizes the interactions between TC internal and external processes for rapid perturbation growth. The success derivation of AI-based CNOPs, with their rapid growth and physical explainability verified in both AI and NWP models, suggests that AI models can capture the most rapidly growing perturbation patterns and their subsequent nonlinear evolution. Thus, potential of AI models is highlighted for advancing atmospheric predictability researches, including theoretical analysis, targeting observations and ensemble forecasts.

# 1 Introduction

The evolution and movement of tropical cyclones (TCs) often trigger multiple disasters, such as gusts, heavy rainfall, and storm surges, posing significant threats to life and property. Therefore, improving the forecast skill of key parameters of TC, particularly the track and landfall position, is vital for mitigating associated risks. Despite substantial progress in numerical weather prediction (NWP), considerable challenges remain in accurately forecasting TC tracks over longer lead times of 3 to 5 days (Yu et al., 2025; Zhou and Toth, 2020). These long-term forecasts are increasingly affected by amplifying multi-scale nonlinear processes (Fovell and Su, 2007; George and Gray, 1976; Holland, 1983), significantly limiting TC track forecast skills in the long run (Lorenz, 1969; Plu, 2011). To address these limitations, advancing our understanding of TC track predictability is imperative. In particular, identifying the dominant sources of forecast errors and elucidating the mechanisms of nonlinear error growth are key research priorities. Insights from such studies are expected to play a critical role in improving targeted observations, refining data assimilation (DA) techniques, and enhancing ensemble forecasting systems, ultimately leading to more reliable long-range TC track predictions (Chen et al., 2025; Lei et al., 2025).

For forecasts of weather phenomena like TCs, predictability is mainly limited by the nonlinear growth of initial-condition-error induced uncertainty. Research approaches aimed at investigating error growth and predictability limits can generally be classified into two main categories, distinguished by the methods used to generate perturbations. The first approach relies on initial perturbations from ensemble-based DA systems. Prominent techniques include ensemble Kalman filter (EnKF; (Evensen, 2003)), as well as its variations (Anderson, 2001; Bishop et al., 2001). These methods aim to generate initial perturbations that statistically approximate the analysis error, and have been widely used to investigate the evolution of initial uncertainties and predictability of TC (Hazelton et al., 2023; Ito and Wu, 2013; Lei et al., 2022; Xu et al., 2025). However, these ensemble-based perturbations represent random samples from the statistical properties of analysis error and thus lack a direct theoretical connection to the dynamical mechanisms underlying error growth in atmospheric systems. Moreover, the reliability of their results is inherently constrained by the specific DA strategies employed and the finite size of the ensemble (He et al., 2020; Liu et al., 2024; Miyoshi et al., 2014)

The second approach to predictability analysis is rooted in dynamical systems theory and focuses on the identification of optimally growing perturbations within atmospheric models. Among the most representative methods is the singular vector (SV), which determines the perturbation vector at the initial time $t_0$ that maximizes linear growth over a finite forecast interval (Buizza and Montani, 1999; Chen et al., 2009). The SV theory and methodology have been extensively implemented in the perturbation generation of the operational ensemble forecasting system at European Centre for Medium-Range Weather Forecasts (ECMWF, 2021). Beyond ensemble forecasting, SVs have been widely applied in studies of atmospheric predictability and targeted observations of high-impact weather events. For instance, (Kim et al., 2011) conducted real-time adaptive observations from dry total energy SVs for TC Jangmi (2008), demonstrating that these dynamics-based SVs effectively captured the underlying mechanisms by which midlatitude troughs and subtropical ridges influence the TC.

Despite substantial progress in the application of the SV methodology, the effective computation of these dynamically-based optimal perturbations continues to depend on a linear approximation (Molteni et al., 1996). This linearization simplifies the complex optimization problem in nonlinear forecast models, thereby enabling its operational implementation. However, the underlying assumption is theoretically valid only for short-term forecasts and sufficiently small perturbations, which limits the applicability of the SV approach to more complex and strongly nonlinear regimes (Mu et al., 2003; Puri et al., 2001).

The Conditional nonlinear optimal perturbation (CNOP) method, originally proposed by (Mu et al., 2003), is designed to identify perturbations that exhibit the maximal nonlinear growth within a finite time interval. As a natural extension of the SV method, CNOP effectively overcomes the linear growth limitation inherent in SVs.

Over the past two decades, CNOP has been extensively applied in predictability studies of weather and climate systems across diverse spatial and temporal scales, including TCs (Mu et al., 2009), the El Nino-Southern Oscillation (Duan and Hu, 2016), Ural blocking events (Li et al., 2024), and Mei-Yu front rainfall (Ke et al., 2022a). Furthermore, CNOP has been successfully employed in real-world targeted observations and ensemble forecasting, demonstrating its potential to enhance the predictive skill of high-impact weather and climate events (Feng et al., 2022; Mu and Duan, 2025; Zhang et al., 2023).

Despite these advantages, CNOP faces challenges during practical application, especially for long-term forecasts. Specifically, extending the optimization time windows to better capture long-term nonlinear error growth substantially increases the computational costs of iterative optimization. In practice, most operational forecast models continue to struggle with accurately and efficiently simulating extreme events such as TCs, thereby constraining the practical implementation of CNOP, particularly in high-resolution models and in real-time, long-range forecasting. Consequently, previous studies have typically adopted the optimization period less than 48 hours when calculating CNOPs for targeted observations and ensemble forecasts of TCs (Niu et al., 2025; Qin et al., 2023; Zhang et al., 2023).

In recent years, Artificial Intelligence (AI) technologies have advanced rapidly, leading to the development of several cutting-edge AI-based weather models, such as Pangu (Bi et al., 2023), GraphCast (Lam et al., 2023), and FuXi (Chen et al., 2023). Leveraging a deep learning (DL) architectures optimized through auto-differential frameworks (e.g. pytorch (Antiga, 2020)) and a graphics processing unit (GPU) acceleration, these models achieve remarkable improvements in both forecast accuracy and computational efficiency compared to the most advanced NWP models. These advantages primarily stem from the highly efficient optimization of DL networks enabled by the embedded auto-differential module. This module can provide convenient backward propagation of gradients, enabling fast solutions of optimization problems. While such gradient-based optimization techniques have been extensively and successfully applied in the training of DL weather models, their potential for solving optimal growth perturbation problems, such as the CNOP, remains largely underexplored. Unlocking this potential could be transformative. Incorporating CNOP into these fast, accurate and gradient-enabled models may enable efficient predictability analysis for long-range forecasts, offering significant potential for further enhancement in forecast skills (Mu and Duan, 2025; Qin et al., 2024).

Despite their promising performance, the physical consistency of AI models is still under debate, raising concerns for the AI-based predictability analysis and applications (Baño-Medina et al., 2025; Bonavita, 2024; Pu et al., 2025; Selz and Craig, 2023). Consequently, the existence and effectiveness of AI-based CNOP have yet to be fully established. (Qin et al., 2024) solved CNOP in AI models for ENSO predictability and demonstrated its growth in numerical models; however, their study focused on climatological mean states rather than individual ENSO events. (Li et al., 2025) applied the CNOP method for 24-hour TC track targeted observations using the FuXi ML-based model, achieving notable improvements in forecast skill. Nonetheless, their analysis did not include a thorough verification of the physical consistency of the derived CNOPs and the sensitive areas. To date, the computation of initial optimal perturbation (i.e., CNOP) that maximizes the nonlinear error growth in extended (5-day) TC track forecast has not been rigorously validated, physically diagnosed, and independently verified. Moreover, the potential impact of the detection of such nonlinear optimal perturbation on the targeted observing design and subsequent forecast skill remains unexplored.

To address this gap, this study develops an optimization framework to derive 5-day CNOP within an AI model, investigates their influence on long-term TC predictability, and evaluates their physical consistency in traditional NWP models. Our work introduces two primary methodological advances. One is the extended optimization window. A five-day optimization window is implemented, facilitating a more comprehensive assessment of the nonlinear processes governing medium-range TC forecasts. The other is the rigorous CNOP validation. A systematic validation of AI-derived CNOPs is performed using both ML-based and traditional NWP models. This significantly broadens the applicability of AI techniques within atmospheric sciences, benefiting fields including theoretical analysis, targeting observation and ensemble forecast.

The remainder of this paper is organized as follows. Section 2 introduces the models, datasets, and TC cases used in this study. Section 3 presents the CNOP methodology and its computational implementation. Section 4 describes the experimental design for CNOP validation and verification. Section 5 reports the CNOP results for different TC cases, while Section 6 further analyzes the sensitivity of TC track errors to initial perturbation regions. Section 7 provides a summary and discussion of the main findings.

## 2 Model and Data

### 2.1 FuXi Machine Learning-based Model

This study employs FuXi, a state-of-the-art AI-based global medium-range weather forecasting model to solve the CNOP problem. According to WeatherBench2 (Rasp et al., 2024), FuXi demonstrates superior performance in RMSE and ACC compared with the ECMWF Integrated Forecasting System (IFS) and other AI-based models.

FuXi consists of three main components: cube embedding, U-Transformer, and fully connected layer. The model is trained with 39 years of data from the fifth generation ECMWF reanalysis (ERA5) data. FuXi predicts a range of meteorological variables, including 5 upper-air atmospheric variables (geopotential, temperature, horizontal wind, and relative humidity) at 13 pressure levels (1000-50hPa) and 5 surface variables (10-m horizontal wind, 2-m temperature, mean sea-level pressure, and 6-hourly total precipitation), with a horizontal resolution of 0.25 degree. This model is initialized using ERA5 reanalysis data, consistent with standard practices employed in other AI-based studies (Baño-Medina et al., 2025; Pu et al., 2025). For comprehensive technical details, readers are referred to (Chen et al., 2023).

## 2.2 Weather Research and Forecast Model

The Weather Research and Forecast (WRF) model, version 4.3 (Skamarock et al., 2019) is utilized in this study to compare the growing dynamics of consistent initial perturbations in AI models and NWP models. As a state-of-the-art NWP system designed for both atmospheric research and operational forecasting applications, WRF is a physics-based numerical model renowned for its capability in simulating and predicting meteorological phenomena across diverse spatiotemporal scales, especially for TC (Zhang et al., 2023) and meso-scale convective systems (Ke et al., 2022b).

In our study, the TC predictability associated with the nonlinear evolution of initial CNOP perturbation in the FuXi model is compared to those in WRF. The integration domain of WRF covers most of the northwestern Pacific Ocean (96.1E-176.1W, 2.0S-60.7N), identical to the initial perturbation region used for the CNOP calculations in FuXi model (see Section 3.2). The WRF model employs a horizontal grid spacing of 15 km which is close to that of the FuXi model for a fair comparison of the results of their error growth dynamics. The vertical hybrid coordinate is discretized into 51 levels, with the top at 30hPa. The integration time step is set to 60 seconds. The model physics parameterization schemes followed the default configuration described by (Skamarock et al., 2019), including Thompson microphysics scheme (Thompson et al., 2008), New Tiedtke cumulus scheme (Zhang and Wang, 2017), the Rapid Radiative Transfer Model for General Circulation Models longwave and shortwave radiation scheme (Iacono et al., 2008), Yonsei University planetary boundary layer scheme (Hong et al., 2006), Noah Land Surface Model (Ek et al., 2003).

## 2.3 TC Case Selection

Twelve representative TC cases exhibiting distinct trajectories, i.e., westward, northwestward, and recurving, between 2020 and 2023 were selected for this study. Each case reached a maximum surface wind speed exceeding 25m/s and exhibited a relatively long lifetimes of more than five days (Figure 1a). The extended duration of these TCs facilitates the computation of the CNOP within a sufficiently long optimization time window and enables a detailed examination of nonlinear perturbation evolution.

For each TC case, the forecast initialization time was selected to ensure that the forecast period encompassed both the stage which the TC maintained an intensity above the tropical storm threshold and the time of its maximum intensity (Table 1). Key information of the selected TC cases is summarized in Table 1. Notably, both the FuXi and WRF models produced relatively accurate simulations of TC tracks across all twelve cases, with mean five-day track errors of 272 km for FuXi and 364 km for WRF, both smaller than the corresponding error of 382 km from the IFS operational forecasts (shown in Figure 1b). In eight of these TC cases, the track errors remain below 200 km across all five lead times (Supplementary Figures 1 and 2). This level of accuracy provides a solid foundation for employing CNOP to investigate the nonlinear error growth in TC track forecasts under relatively realistic conditions. Furthermore, none of the selected TCs were included in the FuXi model's training dataset, ensuring an unbiased comparison of perturbation growth and associated physical processes between the AI-based FuXi model and the physics-based WRF model.

## 2.4 Dataset for TC forecasts and verification

The International Best Track Archive for Climate Stewardship (IBTrACS, (Knapp et al., 2010) was used as the reference dataset for selecting TC cases and evaluating forecast skill. IBTrACS represents a globally coordinated and collaborative effort that integrates TC records from multiple agencies. It provides the most comprehensive global collection of TC best-track data and has been widely adopted in TC researches (Schreck et al., 2014).

## 3 Methodology

## 3.1 Theory of CNOP

The CNOP represents the initial perturbation that induces the largest prediction error within a specified region, under given constraints on the perturbation amplitude and optimization time period. According to the theoretical framework established by (Mu et al. 2003), the CNOP can be formulated as a nonlinear constraint optimization problem:

$$\begin{cases} \max J(\boldsymbol{x_0}) = \|M_t(\boldsymbol{X_0} + \boldsymbol{x_0}) - M_t(\boldsymbol{X_0})\|_{D_1} \\ \quad\quad\quad \|\boldsymbol{x_0}\|_{D_2} \leq \delta \end{cases} \quad (1)$$

where $\boldsymbol{X_0}$ is the unperturbed (or control) atmospheric state at the initial time, $\boldsymbol{x_0}$ is the initial perturbation superimposed on $\boldsymbol{X_0}$. $M_t$ represents the nonlinear propagator of the forecast model at time $t$. $J(\boldsymbol{x_0})$ is the target function measuring the forecast error at time $t$ within the targeted area $D_1$ resulting from the initial perturbation $\boldsymbol{x_0}$, which is to be maximized. $\delta$ is a constant, $\|\boldsymbol{x_0}\|_{D_2} \leq \delta$ gives the constraint on the norm of Initial Perturbations within a specified Area $D_2$ (hereafter called IPA). The optimal initial perturbation $\boldsymbol{x_0}$ solved from the above constrained nonlinear problem is defined as the CNOP.

## 3.2 CNOP Parameter Setting

Before the procedures of calculating the CNOP, the variables and parameters used and relevant physical consideration are introduced first. As we focus on the long-range TC track forecast error, the targeted area $D_1$ is set to center around the TC position in long-range control forecasts at time $t$ (five days in this study). The area $D_1$ is set to an $10° \times 10°$ square area so that the TC inner core and outer rainband are all covered. The norm used for both the targeted function and initial perturbation adopts a typical total dry energy (TDE) norm integrated within a specified region, i.e., the error $\|\cdot\|_D$ is defined as:

$$\|x'\|_D = \frac{1}{2D}\int_D \int_P^0 \left[u'^2 + v'^2 + \frac{c_p}{T_r}T'^2 + R_a T_r \left(\frac{p'_s}{p_r}\right)^2\right] dp \, dD \quad (2)$$

where $D$ is the horizontal area; $p$ is the pressure, serving as the vertical coordinate; $c_p = 1005.7 \text{ J kg}^{-1} \text{ K}^{-1}$ is the specific heat at constant pressure; $R_a = 287.04 \text{ J kg}^{-1} \text{ K}^{-1}$ is the dry air constant; $T_r = 270 \text{ K}$ and $p_r = 1000 \text{ hPa}$ are the reference temperature and pressure. $u'$, $v'$, $T'$, $p_s'$ represent zonal and meridional wind components, temperature, and surface pressure, respectively. This norm reflects errors integrated in different variables on different levels, and has been broadly applied into predictability analysis (Li et al., 2025; Puri et al., 2001). The moisture-related variable is not used because AI models still have difficulty for modelling moist-related processes.

It should be noted that since the TC track is represented as the coordinated positions which is not directly differential, we adopt the TDE as an alternative to approximately measure the TC forecast error. A larger difference (or error) in the TDE within the targeted area $D_1$ is generally related to a significant displacement of TC position. This design of targeted function has been employed in various previous studies (e.g., (Qin and Mu, 2012)).

In this study, we concentrate on TC cases over the Western North Pacific (WNP) Ocean region, the IPA ($D_2$) for the generation of initial perturbations is set to (100-180E, 0-60N), which covers most of the WNP area, and includes main systems affecting TCs, such as subtropical high, midlatitude westly. This domain setup is similar to the design in other Typhoon studies (Guo et al., 2025). For initial perturbations, this study focuses on dry atmospheric processes, and only perturbs temperature and wind above 925hPa at the initial time. The surface variables such as temperature, wind, and surface pressure are excluded since they are strongly influenced by surface boundary conditions like sea surface temperature (SST). This exclusion mitigates potential artificial errors from incomplete representation of cross-system interactions in current AI weather models (Bi et al., 2022; Chen et al., 2023), which ensures reliable initial uncertainty analysis.

For initial perturbation constrain $\delta$ and the optimization periods $t$, six sensitivity analysis are carried out, respectively. The initial perturbation constrain is set to 0.05, 0.1, 0.2, 0.4, 0.8, 1.6 $m^2\,s^{-2}$ in sensitivity analysis (see the test in Section 5.1.3.a) and is finally determined to 0.4 $m^2\,s^{-2}$ in multi-TC experiments for its suitable perturbation strength; The optimization periods are compared among 12, 24, 48, 72, 96, 120 hours (see the test in Section 5.1.3.b) and finally set to 120h in multi-TC experiments for middle-range TC predictivity analysis. Further details in sensitivity analysis can be found in Section 4.

## 3.3 Solving CNOP of TC in FuXi Model

In this section, we give a detailed description of the calculation procedures of CNOP in AI models and its difference from the calculation in numerical models. In our study, CNOP is solved using AI (i.e., FuXi) models with the gradient-based Spectral Projected Gradient 2 (SPG2, (Birgin et al., 2006)) method. This method has been extensively applied in NWP models for solving CNOP. Traditionally in NWP-based studies, tangential and adjoint models for gradient estimation are necessary in gradient estimation, which could consume large amounts of resources and time during their development (Zou, 1997). However, AI models, which are mostly built upon deep-learning frameworks like PyTorch, are equipped with well-developed automatic gradient modules (Antiga, 2020). The gradients of targeted (or loss) functions in AI models are technically available and computationally stable, which greatly reduces the complexity in gradient-based optimization processes. The optimization processes for CNOP using AI-based models include the following four parts corresponding to the schematic diagram in Figure 2:

Step 1. A randomly sampled initial perturbation $x_0$ is projected onto the initial condition $X_0$ and both the control (unperturbed) and perturbed forecasts with a lead of time $t$ (i.e., the optimization period) are produced using the forward AI model $F$. Then the forecast perturbations $x_t$ are given by the difference between the perturbed forecast $F(X_0 + x_0)$ and the unperturbed forecast $F(X_0)$.

Step 2. The target value $J$ corresponding to the forecast perturbation is estimated using the selected norm (i.e., Eq. 2). And its gradient with respect to the forecast perturbation $\partial J/\partial x_t$ is calculated.

Step3. With the advanced auto-differential module, the gradient is propagated backward to the initial perturbations. Thus, the gradient of the target value with respect to the initial perturbation $\partial J/\partial x_0$ is estimated. The effective resolving resolution of a model is about seven times the grid-spacing resolution (about 25 km for FuXi) (Skamarock, 2004), and the scales below this threshold might potentially affect the effective and stable calculation of CNOP. Therefore, scale components of perturbations below 2 degrees (~ 200 km) are removed in gradient.

This step is the critical difference in the calculation of optimal perturbation between AI and numerical models. In numerical models, this step is generally implemented using adjoint model.

Step 4. Based on the estimated gradient, the initial perturbation is updated with the SPG2 algorithm (Birgin et al., 2006). During the optimization, the searching direction is set according to the projected gradient $P\left(x_0 - \alpha \cdot \frac{\partial J}{\partial x_0}\right) - x_0$, where P is the orthogonal projection on the feasible region of the constraint optimization problem (i.e. the constraint of CNOP), and the searching step is determined by linear search. The updated perturbation is used back in Step 1. In this study, the optimization is performed iteratively for 200 times to determine the CNOP. And the increment of target function $J$ is below 0.5 at the end of the optimization.

## 4 Experimental Design

To ensure the applicability of AI-based CNOP in future research on targeted observations and forecasts, particular emphasis is placed on the validation and verification of CNOP in both AI and NWP models. The validation and verification process consist of two complementary steps:

(1) Verification of CNOP growth dynamics in the FuXi model.

Following the methodology of (Ke et al., 2022a), we examine the nonlinear growth dynamics in the FuXi model under different types of initial perturbations. Four types of perturbations are considered: CNOPs, random perturbations (RPs) sampled from a spatially uncorrelated Gaussian distribution, hybrid perturbations composed of CNOPs combined with RPs (CNOP+RPs), and lagged forecast perturbations (LFPs). CNOPs, RPs, and CNOP+RPs are scaled to the same amplitude for direct comparison. LFPs are created using 36h FuXi forecast errors valid at the initial time. More information can be found in supplementary Text S1. Each of these four perturbations is superimposed on identical unperturbed initial conditions to generate perturbed forecasts in the FuXi model. The analysis focuses on comparing the perturbation growth—measured as the differences between perturbed and control forecasts—among CNOPs, RPs, CNOP+RPs, and LFPs. Their values of the targeted function serve as a key diagnostic metric for assessing the optimality of the CNOP solutions.

(2) Verification of physical consistency between CNOPs in FuXi and WRF models.

Previous studies have shown that the skill of AI-based TC track forecasts largely stems from their physically consistent representations of atmospheric circulation evolution (e.g. (Shi et al., 2025)). Thus, the CNOP computed within AI-based models should also represent a fast-growing initial perturbation with physical consistency. Our hypothesis is that applying it as an initial perturbation in a numerical model—specifically, the WRF model—should yield comparable nonlinear spatiotemporal evolution and rapid growth to those observed in the FuXi simulations.

To test the physical consistency of the CNOP between data-driven and numerical models, additional experiments are conducted by introducing these AI-derived CNOP perturbations into the WRF model and perturbing the initial condition. Their growth characteristics are systematically compared with those in FuXi, focusing on the structural evolution and growth rates of the perturbations. This cross-model comparison provides a critical test of the physical consistency and robustness of the CNOP solution derived from the data-driven model.

## 5 Results

A total of twelve TC cases (see Table 1) is utilized for the calculation and validation of CNOP. Before presenting the statistical results for all cases, a single TC case, i.e., Chanthu, is examined in detail to investigate the spatial structure and nonlinear growth characteristics of the CNOP.

### 5.1 CNOP of Typhoon Chanthu

Typhoon Chanthu, characterized by a typical northwestward track followed by a pronounced curvature, is selected for the case study. After reaching tropical storm intensity, Chanthu moves northwestward for approximately 48 hours, before shifting south of Taiwan and subsequently curving northward over the next three days. This trajectory is shaped by complex interplay of multiple weather systems, including the nearby Typhoon Conson to the west, the western boundary of subtropical high, and the mid-latitude westerlies to the north (Figure 3). This subsection presents an analysis of Chanthu's forecast performance, the spatial structure of the CNOP, and the dynamics of perturbation growth.

### 5.1.1 Control Forecast Skills

The nonlinear growth of perturbations relies on the underlying dynamics (or the background flow) of the control forecast. Therefore, it is essential that the control forecasts be sufficiently accurate. Accordingly, we first evaluate the forecast performance of TC track produced by the IFS of ECMWF, FuXi, and WRF, including Chanthu (Figures 3d-3e) and other cases (Figure 1). Direct track comparisons (Figure 3d-e) indicate that Chanthu's track errors simulated by FuXi and WRF remain below 100km throughout the 5-day forecast period, both of which outperform the IFS, especially within the first 96h. The high forecast skill of the FuXi and WRF models in predicting TC tracks is closely associated with their superior performance in simulating TC-related environmental fields, such as temperature and geopotential height (Figures 3a1–a4, 3b1–b4). These simulations exhibit spatial pattern correlations exceeding 0.95 with the ERA5 reanalysis (Figures 3c1–c4) during the first five forecast days.

### 5.1.2 The Structure of CNOP

The CNOP, representing the optimal initial perturbation that induces the maximum nonlinear growth of forecast errors, is computed for Typhoon Chanthu in FuXi. The spatial characteristics of CNOP for different levels and variables are presented in Figure 4. Overall, the CNOP extends across a broad domain, especially at mid- to low-levels from 500 to 850 hPa, encompassing both the TC inner core and remote environmental regions located more than 2000 km away. These broadly-distributed initial perturbations result in a marked westward displacement of the TC track (cf. brown and green curves), shifting the landfall location toward southern China. The underlying mechanisms responsible for this displacement are examined in detail in the following subsection.

The CNOP temperature perturbations are more prominent at the mid- and lower-tropospheric levels than at upper levels. From 925hPa to 500hPa, cold anomalies are concentrated in the southern quadrant within approximately 400 km of the TC center (Figures 4a1-4a3). These anomalies are accompanied by anticyclonic and outward-flow perturbations near the TC core (Figures 4b1-4b3), which act to suppress TC intensification and impede its northward progression. This initial perturbation structure exerts a clear influence on subsequent forecasts. As shown in Figures 5a1 and 5b1, both the FuXi and WRF models simulate the development of a northeast (positive) – southwest (negative) dipole pattern in 500-hPa geopotential height anomalies along the TC track at 24 hours. Typhoon Chanthu tends to propagate across the negative lobe of this dipole in the next four days.

In addition, a distinct warm anomaly is evident approximately 800 km to the northwest of the TC center in the initial CNOP, spanning the northern region of the South China Sea from 925hPa to 500hPa (Figures 4a1-a3). This mesoscale, mid-level warming within the initial perturbation is likely to reduce the pressure at lower levels and enhance the local cyclonic vorticity. Given that TCs tend to migrate along gradients of potential vorticity, such a perturbation may promote a westward displacement of the track. This influence is corroborated in Figures 5a1–5a3, where the perturbed forecast of Chanthu shifts to the northwestward movement toward the South China Sea (brown curve) consistent with the along-track negative anomaly in the evolving 500hPa geopotential height associated with the corresponding initial warm perturbation in CNOP.

For the CNOP in the environmental flow, the initial perturbations are more pronounced at the mid-tropospheric level (around 500hPa) compared to the lower troposphere (Figures 4b1-4b4). Anticyclonic wind perturbations emerge to the north and west of Typhoon Chanthu, located near 20°N, ranging from 120°E to 140°E, approximately 600km away from the TC core. These negative vorticity perturbation act to reinforce the south-western flank of the subtropical high, and to weaken the circulation associated with Typhoon Conson. Their influences are reflected during the forecast in Figure 5 that positive geopotential height anomalies occur near Conson and to the north of Chanthu, contributing to a westward deflection of Chanthu's track. For far-environment perturbations, southerly wind perturbations in the CNOP extend upstream into the mid-latitude trough region over western China (~40°N), far from the TC core (Figure 4b3). These perturbations reduce the northerly flow on the rear side of the trough, leading to its weakening. This suppression of the trough inhibits Chanthu's northward movement, as evidenced by the positive geopotential anomalies over mainland China (Figures 5a1 and 5b1).

### 5.1.3 The Evolution of CNOP

Figure 5 further illustrates the temporal evolution of the CNOP perturbations in both the FuXi and WRF models. As the forecast progresses, the initial perturbations undergo substantial amplification and structural organization. By 24 hours, the perturbations in the vicinity of the TC amplify more rapidly than those in the remote environment, indicating that near-core dipole perturbations play a dominant role in short-range track uncertainty. By 72 hours in the FuXi forecast, the dipole pattern intensifies, with a reconfiguration of mid-latitude long-wave circulation systems. At 120 hours, in both FuXi and WRF, the perturbation structures become more organized. The wave-like 500-hPa geopotential height anomalies at mid-latitude extend from South China to the North Pacific, which is possibly associated with the initial perturbation within the mid-latitude westerly over western China. These changes are dynamically consistent with geostrophic balance; wherein cyclonic wind perturbations align with negative geopotential anomalies and anticyclonic perturbations with positive anomalies.

The optimal growth characteristics of the CNOP are further evaluated in both the FuXi and WRF models, as presented in Figures 5c and 5d. In this assessment, the objective function values associated with the CNOP are compared against those derived from RP and a combination of CNOP and RP. In both models, the CNOP (red stars) consistently yields higher objective function values than either RP (dark green dots), CNOP+RP (light blue dots), or LFPs (orange points), confirming its superiority in capturing the mode of maximal nonlinear error growth. Although the objective function value of the 5-day evolutionary perturbation of the CNOP in WRF is slightly smaller than that in FuXi, owing to differences in their model configurations, it nevertheless represents the maximum among all types of initial perturbations in WRF, thereby confirming the physical optimality of the CNOP. Furthermore, the evolutions of TC track errors induced by CNOP perturbations are broadly consistent in both models, with track deviations exceeding 1400 km by the 5-day forecast (Figure 5e). This physical consistency highlights the robustness and validity of the CNOP methodology when computed using AI-based systems, and further strengthens confidence in the CNOP approach for identifying dynamically relevant sources of forecast uncertainty.

### 5.1.4 Sensitivity Analysis

Based on the CNOP solution and its verification within TC Chanthu, a parameter sensitivity analysis is conducted to investigate the factors influencing the CNOP results. In particular, the sensitivity of CNOP to two key parameters—the constraint magnitude ($\delta$) and the optimization time window ($t$) (see their definitions in Section 3.2)—is examined.

### 5.1.4.a Sensitivity to magnitude constraint

In the case study of Chanthu above, the CNOP is derived given the norm constraint of TDE of 0.4 m² s⁻², which gives perturbation with magnitude approximate to the analysis error estimation (Feng et al., 2023; Leutbecher and Palmer, 2008). To investigate the sensitivity of the CNOP to the magnitude constraint, we conducted additional experiments by tuning the constraint to 0.05, 0.1, 0.2, 0.8, and 1.6 $m^2\ s^{-2}$. The optimization time window is fixed at 120 hours to emphasize long-term predictability.

Figure 6 presents the CNOPs obtained under the constraints of 0.1, 0.8, and 1.6 $m^2\ s^{-2}$, along with their corresponding TC tracks from the perturbed forecasts. It is apparent that larger constraint magnitudes produce perturbations of greater amplitude and broader spatial extent. Specifically, for a smaller constraint (e.g., 0.1 $m^2\ s^{-2}$), the perturbations are primarily confined to the tropical region surrounding the TC vortex within a radius of approximately 600 km. This CNOP configuration induces a relatively small deviation in the TC track—less than 400 km at five days.

When the magnitude constraint increases to 0.8 $m^2\ s^{-2}$, the CNOP perturbations extend into the mid-latitude (~30°N) regions of western China. As a result, in contrast to the small magnitude constraint, this wider-spread CNOP produces a remarkably diverged TC track which ultimately makes landfall in south China—approximately 1600 km away from the control track. Under the largest constraint of 1.6 $m^2\ s^{-2}$, the perturbations intensify further in the mid-latitudes and even extend to the vicinity of the cutoff low near 45°N. The comparison clearly demonstrates that with the increase of the magnitude constraint, the optimal initial perturbation (CNOP) tends to extend over broader and more distant regions within the mid-latitude baroclinic environment. This behavior arises because synoptic-scale baroclinic systems exhibit larger spatial variability than smaller-scale vortex structures (Judt, 2020). When a larger magnitude constraint is applied, perturbing the remote synoptic-scale environment becomes more effective in inducing substantial divergence in TC tracks. These findings underscore the CNOP algorithm's capability to automatically identify physically meaningful sensitive regions of initial perturbation, provided that an appropriate magnitude constraint is specified.

Similar as Figure 5, analysis on perturbed TC forecasts and CNOP verifications under different constraints are presented in Figure 7. The error evolution indicates that larger perturbation constraints result in greater westward shifts of TC tracks, accompanied by greater track forecast errors and larger target values in both FuXi and

WRF. More importantly, the target values of CNOPs obtained under various constraints remain higher than those derived from RPs and CNOP+RPs in both FuXi and WRF, suggesting strong physical consistency of CNOP from FuXi across different constraint magnitudes. It is also confirmed by high spatial similarity between the perturbation growth of CNOPs in FuXi and WRF, as shown in Supplementary Figures 3 and 4.

### 5.1.4.b Sensitivity to optimization periods

In addition to examining the magnitude constraint of the CNOP, its dependence on the optimization period is also investigated. The optimization periods are set to 12, 24, 48, 72, 96, and 120 hours, respectively, with the initial constraint fixed at 0.4 $m^2\ s^{-2}$ as determined from previous analyses.

Figure 8 presents the CNOPs optimized for different periods. Although all perturbations satisfy the same constraint of 0.4 $m^2\ s^{-2}$, those optimized over longer periods exhibit broader spatial coverage. Specifically, the CNOP optimized for a 24-hour period shows perturbations concentrated near the TC center, where wind perturbations can exceed 2 $m\ s^{-1}$. These strong initial perturbations near the TC vortex induce a noticeable northward displacement of the TC within the first 48 hours, but this influence diminishes at longer lead times, resulting in relatively close trajectories between the control and perturbed experiments.

In contrast, the CNOP optimized for a 5-day period exhibits perturbations distributed over a much wider area, extending from the western Pacific to western China and affecting multiple weather systems. Such far-field perturbations exert a more sustained influence on TC motion, leading to a more pronounced westward bias in the TC track (see Figure 4). In 3-day optimized CNOP, perturbations are mainly in tropical area, and the westward bias only last for the first 72h. After that, TC undergoes a northward movement. This finding suggests that perturbations in the TC's remote environment become increasingly important for forecast uncertainty at extended lead times.

Figure 9 illustrates the CNOP-perturbed TC forecasts, CNOP-induced error growth, and target function values in both the WRF and FuXi models. For track errors, each CNOP yields its maximum deviation at the corresponding optimization period, underscoring the validity and rationality of defining CNOPs based on specific forecast horizons. The target function values of CNOPs increase with longer optimization periods in both models and exceed those derived from RP and CNOP+RP methods, demonstrating the superior optimality of CNOPs in both FuXi and WRF. Moreover, the evolution of 500-hPa geopotential height errors, as shown in Supplementary Figures 5 and 6, exhibits high consistency between the two models, indicating the physical realism of AI-based CNOPs.

## 5.2 Multi-TC Predictability Analysis

To assess the overall validity of CNOP computation in data-driven models, 12 TC cases that occurred over the northwestern Pacific during 2020–2023 were examined. Consistent with the experiment for Typhoon Chanthu, the initial perturbation constraint was set to 0.4 $m^2\ s^{-2}$, ensuring an appropriate perturbation amplitude, while the optimization period was fixed at 120 hours to facilitate long-term predictability analysis.

### 5.2.1 Inner vs. environmental perturbations of CNOPs

The CNOPs for all 12 TC cases were calculated using the FuXi model, with the results illustrated for temperature (shading) and wind (vectors) at 850 hPa (Figure 10) and 500 hPa (Supplementary Figure 7). It is evident that the optimally growing initial perturbations (i.e., CNOPs) exhibit pronounced case dependence, closely associated with the prevailing weather regimes. Furthermore, the spatial structures of the CNOPs in most TC cases extend beyond the near-vortex region, encompassing remote environmental areas as well. This finding indicates that the amplification of track errors arises from the combined effects of error growth within both the inner-core and environmental regions—an interaction effectively captured by the CNOP approach. Nonetheless, the relative magnitudes of the initial perturbations in the near-vortex and environmental regions vary substantially across different cases. To further investigate this intriguing behavior, two representative TCs—Chanhom and Maysak—were selected for detailed comparative analysis.

The CNOP structures of TCs Chanhom and Maysak at 850 hPa are shown in Figures 10b and 10a, respectively. The remarkable difference is that the CNOP of TC Maysak encompasses not only the vortex itself but also the near and far environments, while the CNOP structure of TC Chanhom is concentrated primarily around the TC vortex, as also shown for 500 hPa (Supplementary Figure 7). In particular, the vortex perturbations associated with Maysak exhibit magnitudes comparable to those observed in both the near and far environments—even beyond 2000 km—standing in stark contrast to the predominantly near-vortex perturbations characteristic of Chan-hom. These findings are consistent with the results of (Feng et al., 2024).

Specifically, at 850hPa, TC Maysak exhibits large perturbations spanning from the southern Philippines to northern Japan, with extensions into the upstream ridge region of the mid-latitudes. Cold anomalies near the TC center weaken the cyclone intensity and accelerate its movement, while warm anomalies and wind perturbations in the periphery modify the surrounding synoptic systems. In TC Chanhom, however, perturbations are mainly confined near the TC core. At 850 hPa, cold anomalies and eastward wind perturbations appear near the TC center, promoting eastward displacement, whereas warm anomalies and anticyclonic wind perturbations to the southwest enhance the adjacent subtropical high. As a result, the perturbed TC tracks display an evident eastward bias.

## 5.2.2 The CNOP evolution for multiple cases

The evolution of CNOPs for all 12 TC cases was further analyzed using both FuXi and WRF models, with the results presented in Supplementary Figures 8–11. In the main text, the temporal evolution of CNOPs for TCs Chanhom and Maysak is discussed in detail, following the analysis described in the previous section.

Figures 11 and 12 depict the perturbation growth of TCs Chanhom and Maysak in both FuXi and WRF. In both cases, perturbations primarily intensify near the TC center and within the mid-latitude region, though the impact on the TC track is more pronounced for Chanhom (Figure 12). In Chanhom, perturbations originating in the core region rapidly propagate into the mid-latitude westerlies. The amplification of these perturbations modifies the westerly trough structure, resulting in a notable eastward displacement of the TC track. In contrast, Typhoon Maysak is situated far from the westerly trough and lacking adjacent strong synoptic systems. Its environmental perturbations primarily evolve through their own nonlinear development associated with baroclinic instabilities, with minimal feedback from the TC's internal dynamics. This comparison indicates that CNOP evolution effectively captures the key physical processes responsible for TC track forecast sensitivity in different TC cases. A detailed comparison of the relative contributions of core, mid-range, and peripheral perturbations for these two cases is provided in Section 6.

For the remaining TC cases (Supplementary Figures 8–11), perturbation growth also occurs predominantly near the TC centers and within the mid-latitude regions. The consistent results obtained from FuXi and WRF further confirm the physical reliability and robustness of AI-based CNOPs in representing perturbation evolution across multiple modeling systems.

To further assess the optimality and physical consistency of CNOPs, the target function values of CNOPs, RPs, CNOP+RPs, and LFPs obtained from FuXi and WRF are compared in Figures 13a–b, while the CNOP-induced TC track errors from both models are summarized in Figures 13c–d. The corresponding results for individual TC cases are presented in Supplementary Figures 12 and 13. Consistent with the findings for TC Chanthu, both models show that the target values derived from CNOPs are consistently larger than those from RP, CNOP+RP, or LFPs, demonstrating the superior optimality of the AI-based CNOP approach in both the AI-driven FuXi model and WRF.

## 6 Sensitivity of TC track error to initial perturbation regions

The preceding analyses have verified both the physical consistency and optimality of AI-based CNOPs. In this section, to further examine the sensitivity of TC tracks to internal and environmental perturbations, the influences of three distinct regions—the TC inner core (a 10° × 10° area centered on the TC), the near-TC environment (a 20° × 20° area centered on the TC, excluding the inner core), and the far-field environment (the remaining domain beyond the middle region) — on TC track uncertainties are compared. This experiment is conducted using three representative TC cases: Chanhom, Maysak, and Chanthu, consistent with the cases analyzed in the main text. Their extracted perturbations are shown in Supplementary Figure 14, and their respective impacts on TC track forecasts are presented in Figure 14.

As shown in Figure 14, all three TC cases exhibit the greatest overall sensitivity to CNOP perturbations, though the degree of sensitivity to different CNOP regions varies among cases. Generally, regions with larger CNOP magnitudes exert greater influence. For instance, not all TCs are most sensitive to environmental perturbations. In the case of Chanhom, whose CNOP is concentrated near the TC core, the internal perturbations induce the largest track deviations. Conversely, in Maysak and Chanthu, where CNOPs are broadly distributed across the environmental field (see Figures 4 and 10), the outlying perturbations have the most pronounced impact on TC track forecasts. These findings align with those of (Feng et al., 2024) and highlight the case-dependent nature of TC sensitivity.

Such variability implies that the dominant sensitivity regions differ among TCs and should be dynamically identified through the CNOP framework in operational forecasting. A detailed discussion of the sensitivity regions derived from AI-based CNOPs and their applications to targeted observations is provided in following study. Furthermore, the track errors produced by full CNOP perturbations are substantially larger than those resulting from perturbations in individual regions, underscoring the importance of the interactions between internal and environmental components. This finding further demonstrates the advantage of the CNOP approach in capturing the coupled processes that govern TC predictability.

# 7 Conclusion and discussion

In predictability analysis, the optimally growing perturbation patterns, such as the SV and the CNOP in linear and nonlinear frameworks, respectively, can lead to great uncertainties in forecasts. These perturbations are of fundamental importance for both theoretical studies and practical applications in weather and climate prediction. Traditionally, their computation and analysis have relied on NWP systems. However, the limited computational efficiency of these models poses significant challenges to long-term predictability analysis. Recent advances in AI have opened new avenues for rapid and accurate forecasting, extending to prediction horizons of up to 15 days. Leveraging automatic differentiation and the GPU acceleration within deep-learning frameworks, AI models can efficiently generate optimally growing perturbations, significantly advancing the long-term predictability analyses and applications.

This study develops an optimization framework for deriving the Conditional Nonlinear Optimal Perturbation (CNOP) for long-term (5 days) TC predictions within the framework of the AI-based FuXi model. The nonlinear optimality and physical explainability of the CNOP is rigorously examined and validated in both AI and NWP frameworks. Specifically, this study systematically investigates the sensitivity of CNOP to variations in the constraint on initial perturbation magnitudes and optimization periods. In addition, the target function, spatial structures, and nonlinear evolutions of the CNOPs are comprehensively analyzed using twelve representative TC cases. To our knowledge, no previous studies have conducted such a thorough analysis and verification of optimally growing perturbations in the context of long-term TC forecasting.

Firstly, the 5-day optimized CNOP is analyzed for Typhoon Chanthu. The CNOP perturbations are primarily concentrated in regions that exert significant influence on TC forecasts, including mid-latitude waves, the subtropical high, and nearby tropical cyclones. These perturbations alter the typhoon's trajectory from a northward to a westward motion, resulting in a track deviation exceeding 1,400 km in both the FuXi and WRF models. Beyond the track error, the evolution of perturbations in the environmental fields exhibits strong consistency between FuXi and WRF. In both models, the geopotential height perturbations amplify rapidly and organize into a large-scale, wave-like structure extending from the tropics to the mid-latitudes. The optimality of the CNOP is further verified in both models, as its objective function value surpasses those of other perturbations, demonstrating its nonlinear optimality and physical consistency.

Building on the optimal perturbation obtained for Typhoon Chanthu, the parameter sensitivity of the CNOP is further investigated. For different initial perturbation magnitudes, stronger perturbations are found to extend over broader spatial regions, exerting greater influence on forecast outcomes. The spatial discrepancies among CNOPs constrained by varying magnitudes highlight the inherently nonlinear nature of the error growth mechanism. When varying the optimization period, CNOPs optimized over longer time intervals encompass larger spatial scales and incorporate a wider range of dynamical processes relevant to long-term TC forecasts. Importantly, the growth characteristics of these CNOPs exhibit strong agreement between the FuXi and WRF models. Their optimality is consistently confirmed in both frameworks, as evidenced by superior objective function values compared to other perturbation types.

This study extends the CNOP analysis to twelve distinct TC cases. The results demonstrate that AI-based CNOPs can generate rapidly growing and physically consistent initial perturbations that exert a pronounced influence on TC track forecasts. Across different cases, the spatial structures of CNOPs vary considerably, reflecting the diverse optimal dynamical processes associated with each TC. Verification of their optimality in both the FuXi and WRF models confirms the physical consistency of AI-derived CNOPs across a wide range of TC events. Further examination of the relative contributions of different CNOP components reveals strong case dependence: not all TCs exhibit the greatest sensitivity to environmental steering flows, consistent with the findings of (Feng et al., 2024). Moreover, the system interactions captured by CNOPs are shown to play a crucial role in facilitating rapid perturbation growth. These findings highlight the necessity of case-specific CNOP optimization for improving the accuracy of predictability assessments.

In summary, this study demonstrates that perturbations with specific spatial structure exhibit significant development in both AI and NWP models. The physical consistency of AI models enables them for a wide range of applications, including operational forecasting, predictability analysis, and dynamical mechanism exploration. These findings represent an important extension of AI's utility in atmospheric sciences. For instance, in following study, the AI model will be employed to identify sensitive regions for targeted observations, and the potential for enhancing practical forecast skill through the assimilation of additional observations in these regions will be evaluated—advancing the transition from "big data" to "better data."

While the present study primarily focuses on the role of initial condition uncertainties in TC track forecasts, future research should also address uncertainties inherent in AI models themselves, as these constitute a major source of forecast error. Moreover, extending the analysis to TC intensity predictability, which is equally critical for operational forecasting, will be a key direction for further investigation. Continued AI-based predictability studies will provide deeper insights into the physical and dynamical characteristics of these models, ultimately supporting their refinement and broader integration into operational forecasting systems.

## Funding Statement

This study is supported by the National Natural Science Foundation of China (42288101) and the Academician Workstation of AP-TCRC. The computations in this research were performed using the CFFF platform of Fudan University.

## Data Availability Statement

ERA5 data is available at https://cds.climate.copernicus.eu/cdsapp#!/search?type=dataset; IBTrACS dataset is at

https://www.ncei.noaa.gov/products/international-best-track-archive; FuXi model is open sourced at https://github.com/tpys/FuXi; WRF model is available at https://www2.mmm.ucar.edu/wrf/users/download/get_sources_new.php

## Conflict of Interest Statement

The authors declare there are no conflicts of interest for this manuscript.

## Acknowledgments

The authors would like to thank Drs. Bo Qin and Guokun Dai for their valuable comments on this study.

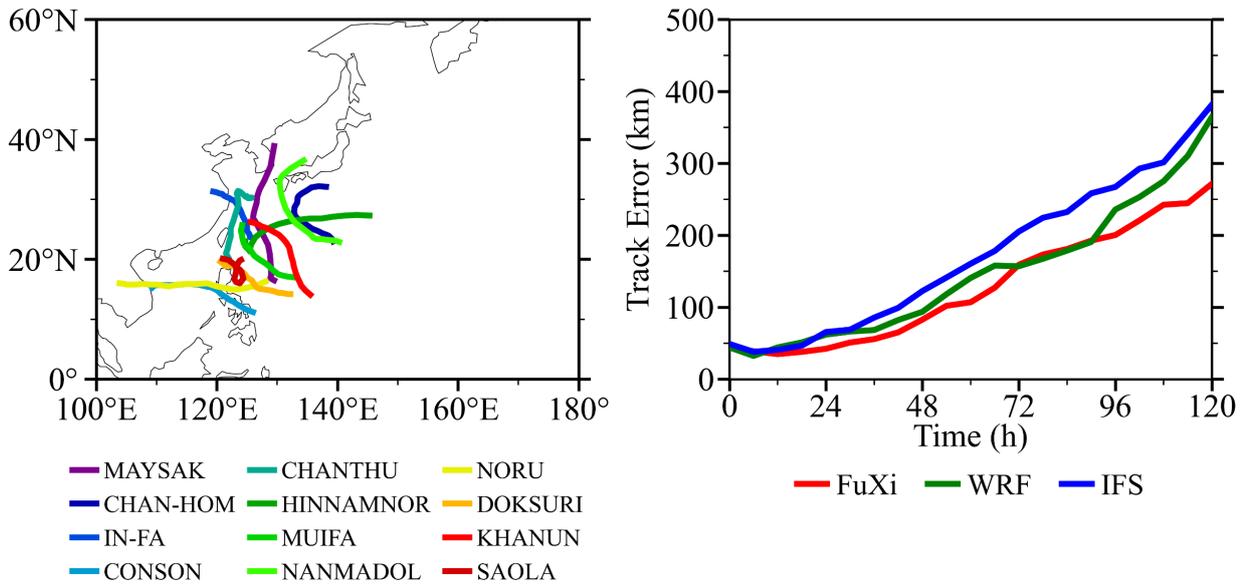

**Figure 1.** Tracks and forecast errors of selected TC cases. Figure (a) gives TC tracks from IBTrACS dataset. Figure (b) gives mean track errors from FuXi, WRF and IFS.

**Table 1.** Tropical cyclone cases

| TC cases | Initial Time | Maximum Intensity and Time | Duration |
|---|---|---|---|
| Maysak | 0000 UTC Aug 29, 2020 | 61.7m/s, 930hPa, 0000 UTC Sep 01, 2020 | 1200 UTC Aug 26-0000 UTC Sep 7, 2020 |
| Chan-hom | 0000 UTC Oct 05, 2020 | 41.2m/s, 963hPa, 1800 UTC Oct 07, 2020 | 0600 UTC Oct 03-0600 UTC Oct 19, 2020 |
| In-Fa | 0000 UTC Jul 22, 2021 | 46.3m/s, 954hPa, 0000 UTC 22 Jul, 2021 | 1800 UTC Jul 15-0600 UTC 31 Jul 2021 |
| Conson | 1200 UTC Sep 06, 2021 | 30.9m/s, 980hPa, 1200 UTC Sep 06, 2021 | 0000 UTC Sep 05-1200 UTC 13 Sep, 2021 |
| Chanthu | 0000 UTC Sep 11, 2021 | 74.6m/s, 915hPa, 0000 UTC Sep 11, 2021 | 0600 UTC Sep 05-0000 UTC Sep 20, 2021 |
| Hinnamnor | 0000 UTC Aug 29, 2022 | 72.0m/s, 915hPa, 0000 UTC Sep 01, 2022 | 0000 UTC Aug 27-0000 UTC Sep 09, 2022 |
| Muifa | 0000 UTC Sep 08, 2022 | 56.6m/s, 946hPa, 0600 UTC Sep 11, 2022 | 1800 UTC Sep 03-0000 UTC Sep 17, 2022 |
| Nanmadol | 1200 UTC Sep 14, 2022 | 69.5m/s, 916hPa, 1800 UTC Sep 16, 2022 | 0600 UTC Sep 11-0600 UTC Sep 20, 2022 |
| Noru | 0000 UTC Sep 24, 2022 | 74.6m/s, 914hPa, 1800 UTC Sep 24, 2022 | 0000 UTC Sep 21-0600 UTC Sep 29, 2022 |
| Doksuri | 1200 UTC Jul 21, 2023 | 66.9m/s, 928hPa, 0000 UTC Jul 25, 2023 | 0600 UTC Jul 07-0000 UTC Jul 31, 2023 |
| Khanun | 1200 UTC Jul 28, 2023 | 64.3m/s, 924hPa, 0600 UTC Aug 01, 2023 | 0600 UTC Jul 27-1800 UTC Aug 11, 2023 |
| Saola | 0000 UTC Aug 25, 2023 | 72.0m/s, 917m/s, 1800 UTC Aug 29, 2023 | 0000 UTC Aug 22-0000 UTC Sep 04, 2023 |

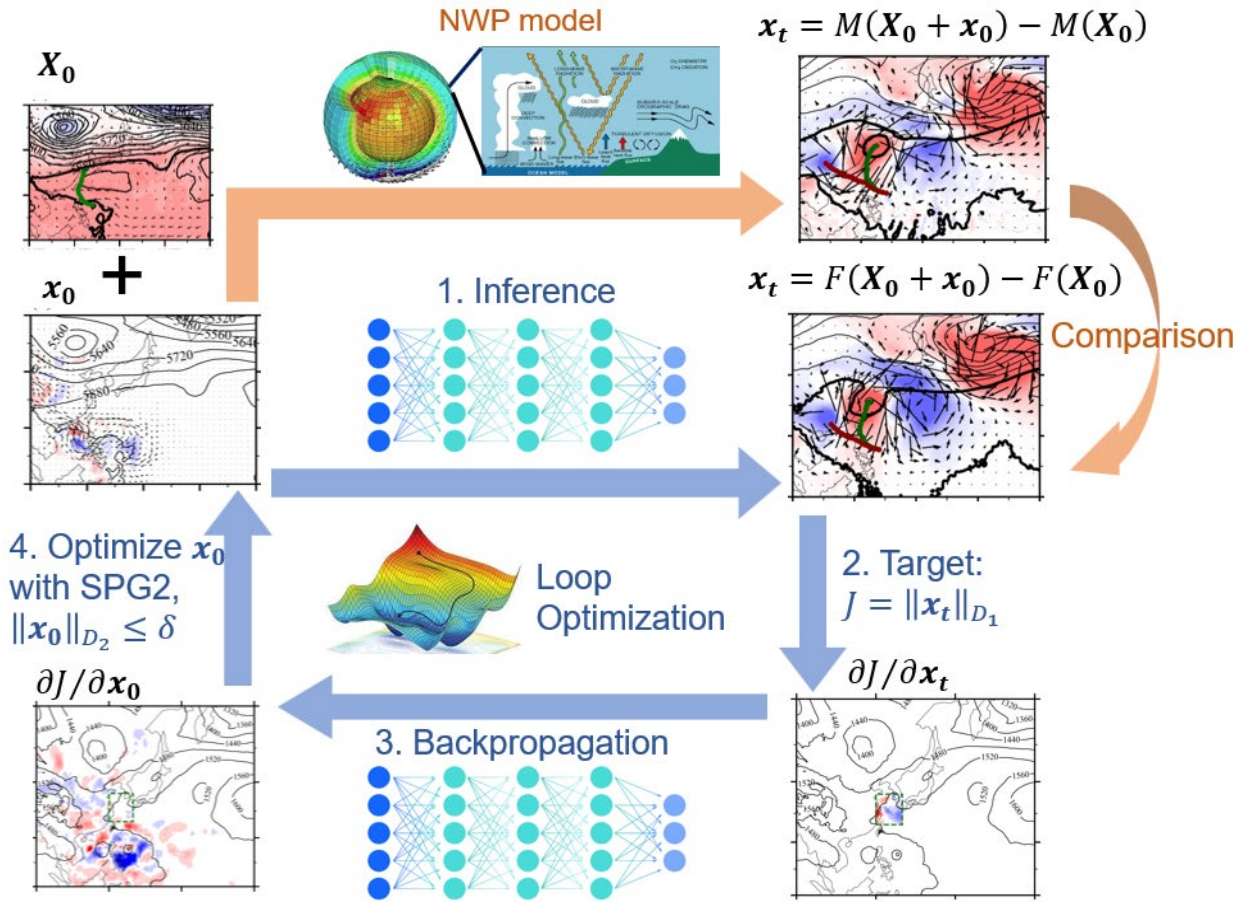

**Figure 2.** The procedures for solving CNOP in data driven model (blue arrows) and for verifying CNOP in NWP models.

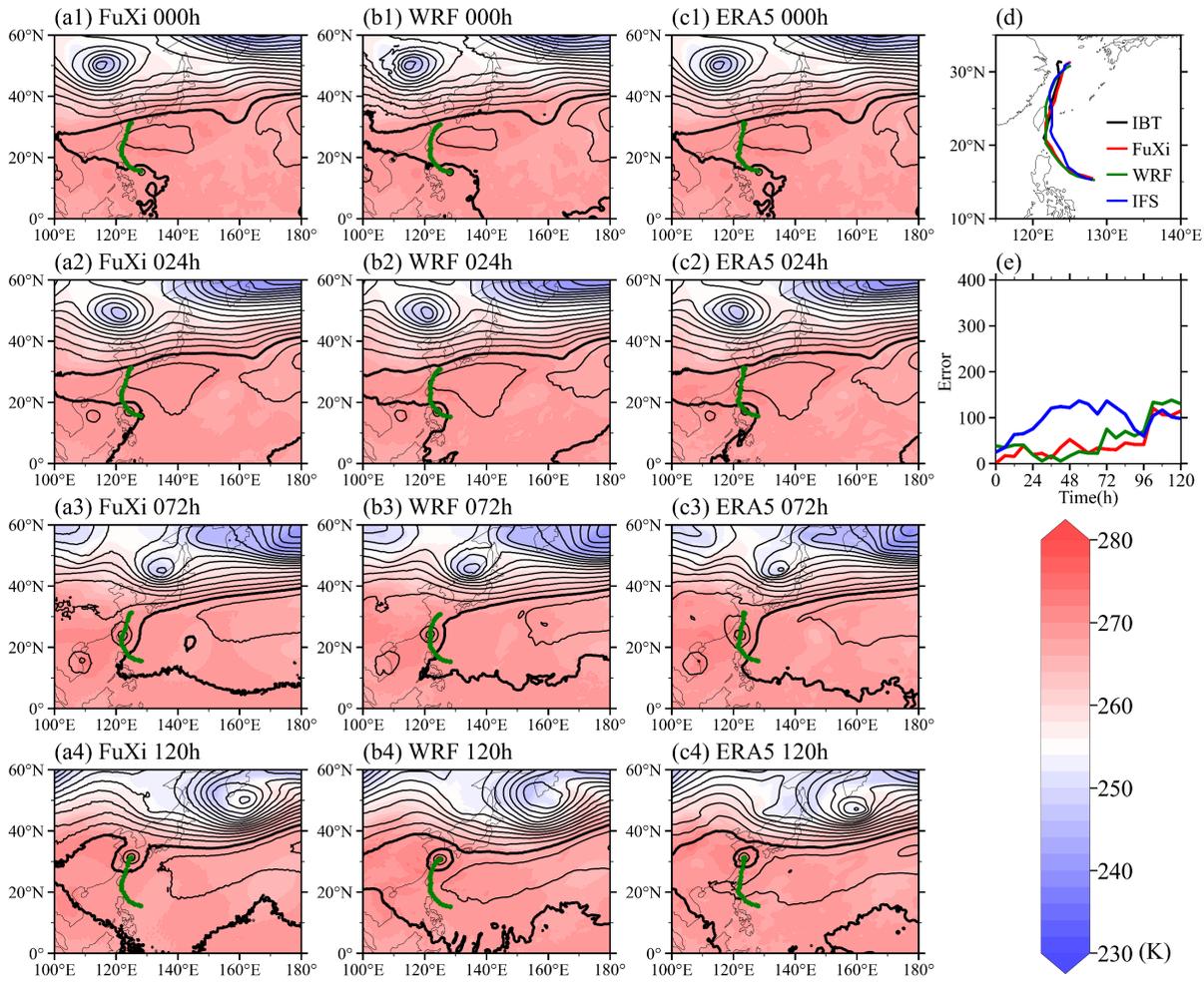

**Figure 3.** The 120h forecast and analysis of TC Chanthu initialized at 0000 UTC Sep 9, 2021. Figures (a1-a4) give the 0h, 24h, 72h, 120h forecast of 500hPa geopotential field (contours, interval at 40gpm, thickened at 5880gpm) and temperature field (shadings). Green lines for TC track. Figures (b1-b4) and (c1-c4) are the same as (a1-a4), but for results from WRF and reanalysis from ERA5. Figure (d) compares the TC track forecasts from FuXi (red), WRF (green), IFS (blue) and track analysis from IBTrACS (black). Figure (e) compares the track errors from different forecasts.

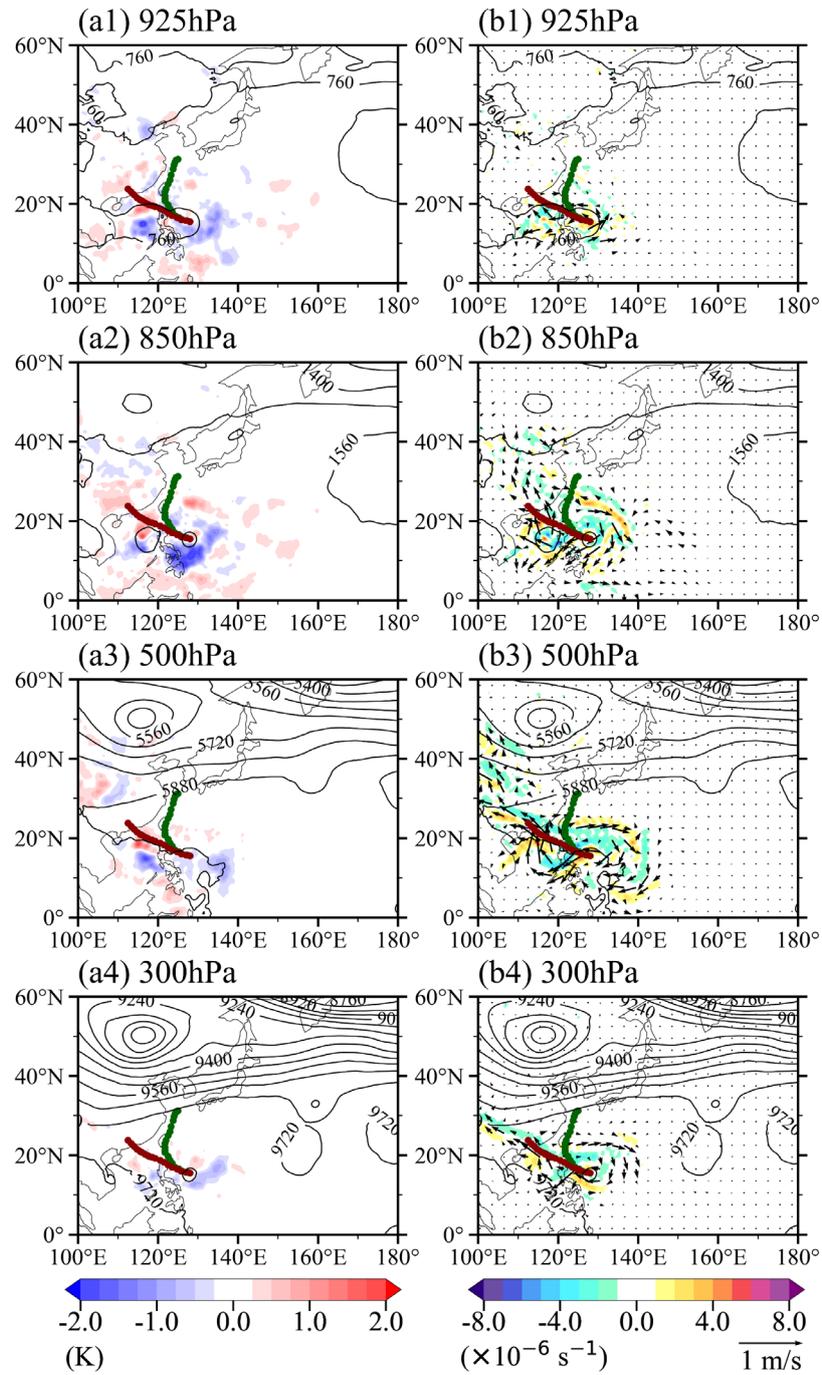

**Figure 4.** CNOP structures of TC Chanthu. Figures (a1-a4) gives initial temperature perturbations (shadings) on isobaric surface of 925hPa, 850hPa, 500hPa and 300hPa, respectively. Figures (b1-b4) are similar as (a1-a4), but for initial wind perturbations (vectors) and its corresponding vorticity field (shadings). Contours represent the geopotential field. Green lines and red lines give TC tracks from control forecasts and perturbed forecasts.

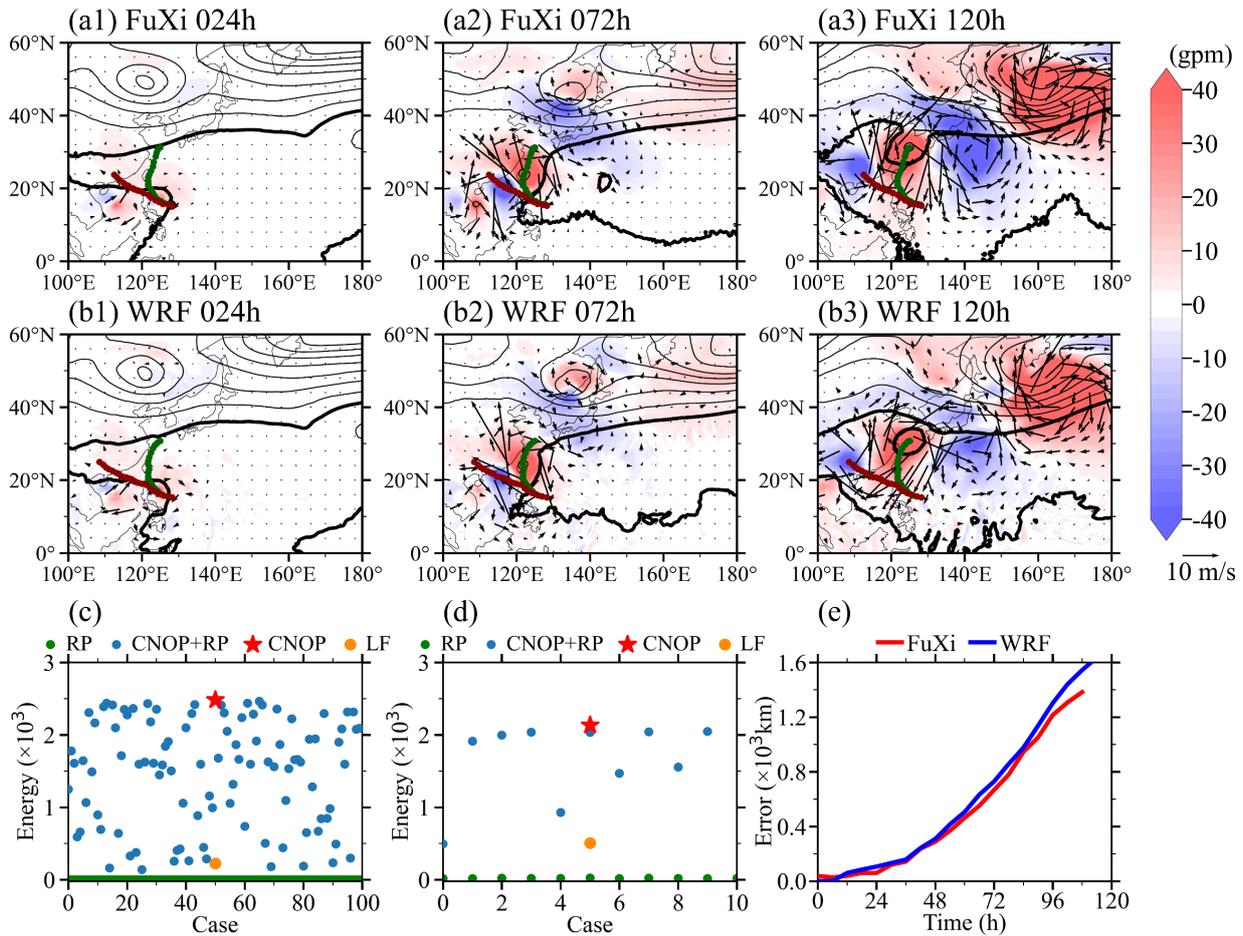

**Figure 5.** The evolution and verification of CNOP for TC Chanthu. Figures (a1-a3) gives the 24h, 72h, 120h evolution of 500hPa geopotential perturbation (shading) and wind perturbation (vectors) in FuXi. Contours give the geopotential height from control forecast (interval at 80gpm and thickened at 5880gpm). Green lines and red lines represent TC tracks from control forecast and perturbed forecast. Figures (b1-b3) are similar as (a1-a3) but for perturbation evolutions in WRF. Figure (c) gives the target values of CNOP (red star), random perturbation (RP, green dots), lagged forecast perturbations (LFPs, orange dots) and CNOP combined with RP (blue dots) from FuXi. Figure (d) is similar as Figure (c), but the target values are evaluated with WRF. Figure (e) gives the TC track error caused by CNOP in FuXi (red line) and WRF (blue line).

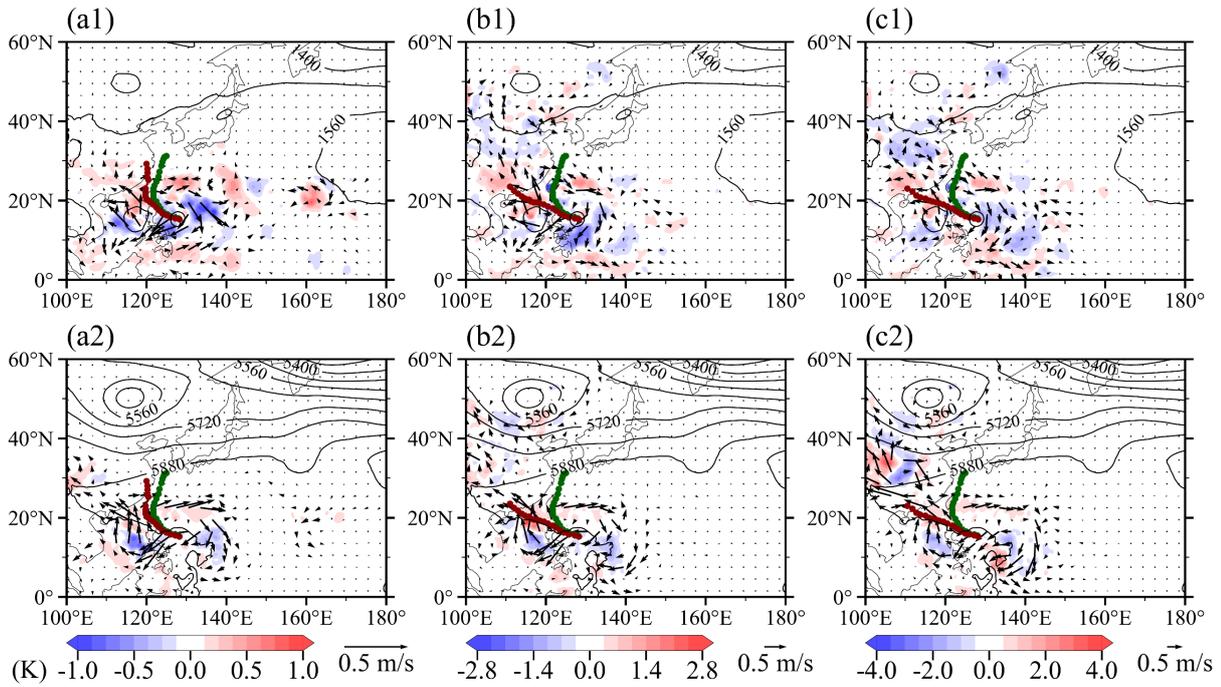

**Figure 6.** CNOP structures of TC Chanthu with different initial norms. Figure (a1, a2) gives initial perturbations of temperature (shading) and wind (vectors) on 850hPa and 500hPa with an initial norm of 0.1. Contours represent geopotential heights. Green lines and red lines represent TC tracks from control forecast and perturbed forecast in FuXi. Figures (b1, b2) and (c1, c2) are similar as Figures (a1, a2), but for initial norms of 0.8 and 1.6.

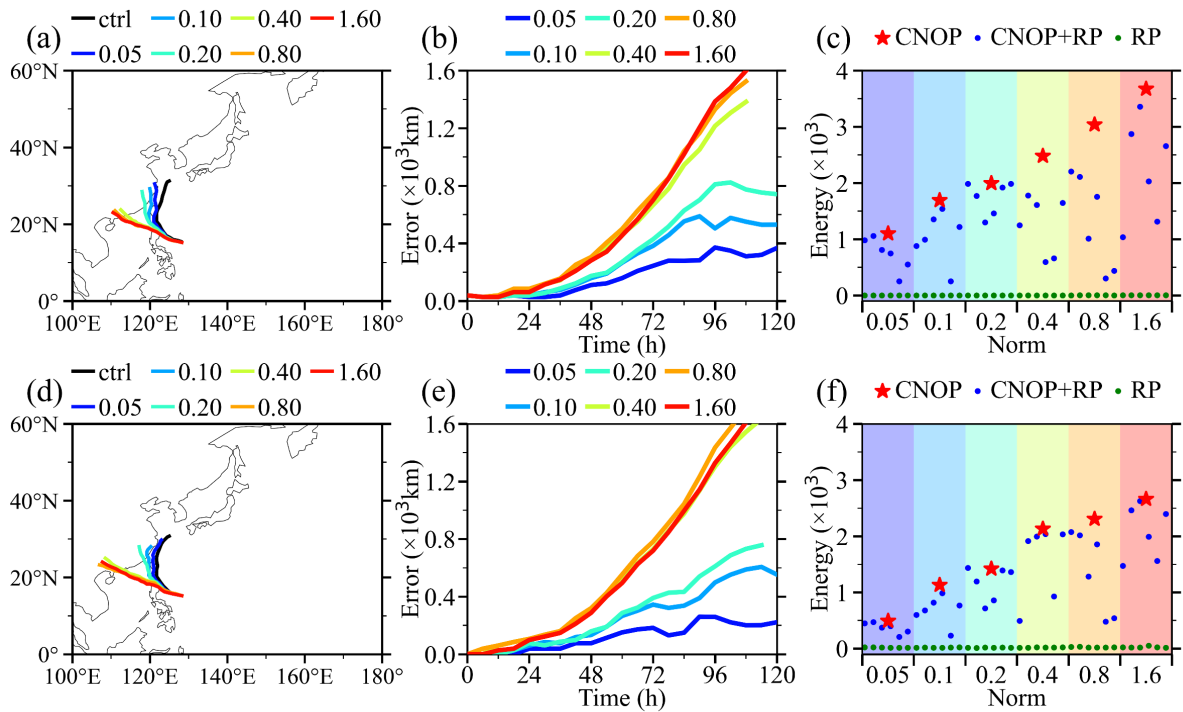

**Figure 7.** Forecasts of TC Chanthu perturbed by CNOPs with different norms. Figure (a) gives TC tracks of control and CNOP-perturbed forecasts from FuXi. Figure (b) gives track errors caused by different CNOPs in FuXi. Figure (c) gives target values of CNOPs with different norms (red star), random perturbation (RP, green dots) and CNOPs combined with RPs (blue dots) evaluated from FuXi. Figures (d-f) are similar as Figure (a-c), but for results from WRF.

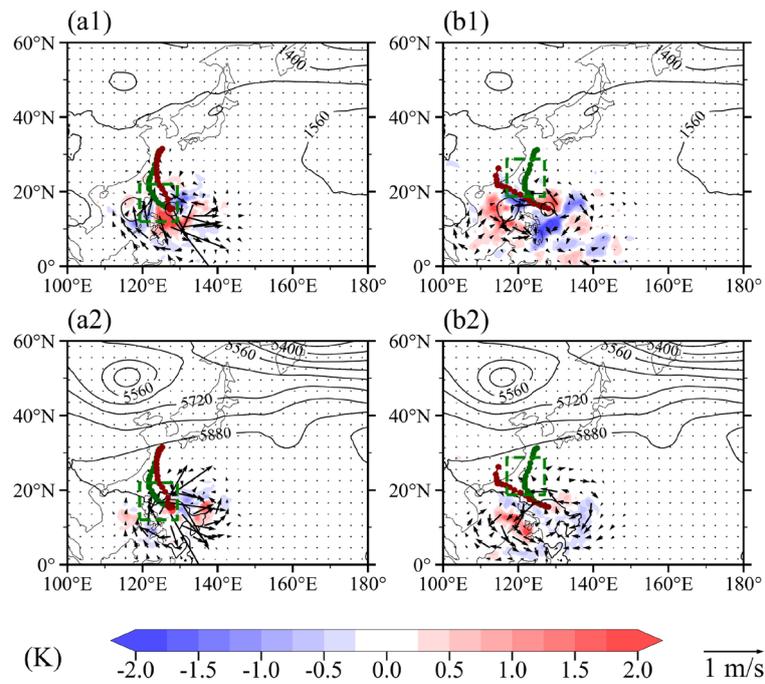

**Figure 8.** CNOP structures of TC Chanthu with different optimization times. Figures (a1, a2) gives initial perturbations of temperature (shading) and wind (vectors) on 850hPa and 500hPa with an optimization time of 24h. Contours represent geopotential heights. Green lines and red lines represent TC tracks from control forecast and perturbed forecast in FuXi. Figures (b1, b2) are similar as Figures (a1, a2), but for optimization times of 72h.

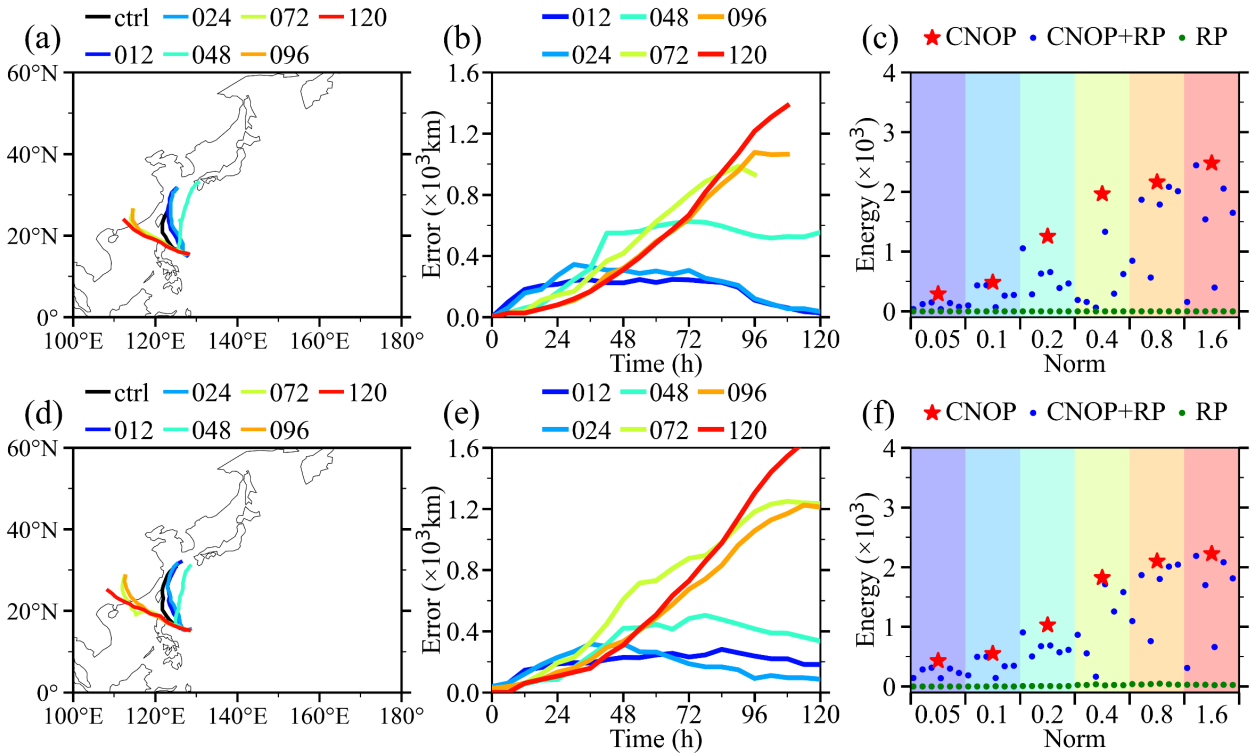

**Figure 9.** Forecasts of TC Chanthu perturbed by CNOPs with different optimization times. Figure (a) gives TC tracks of control and CNOP-perturbed forecasts from FuXi. Figure (b) gives track errors caused by different CNOPs in FuXi. Figure (c) gives target values of CNOPs with different optimization times (red star), random perturbation (RP, green dots) and CNOPs combined with RPs (blue dots) evaluated from FuXi. Figures (d-f) are similar as (a-c), but for results from WRF.

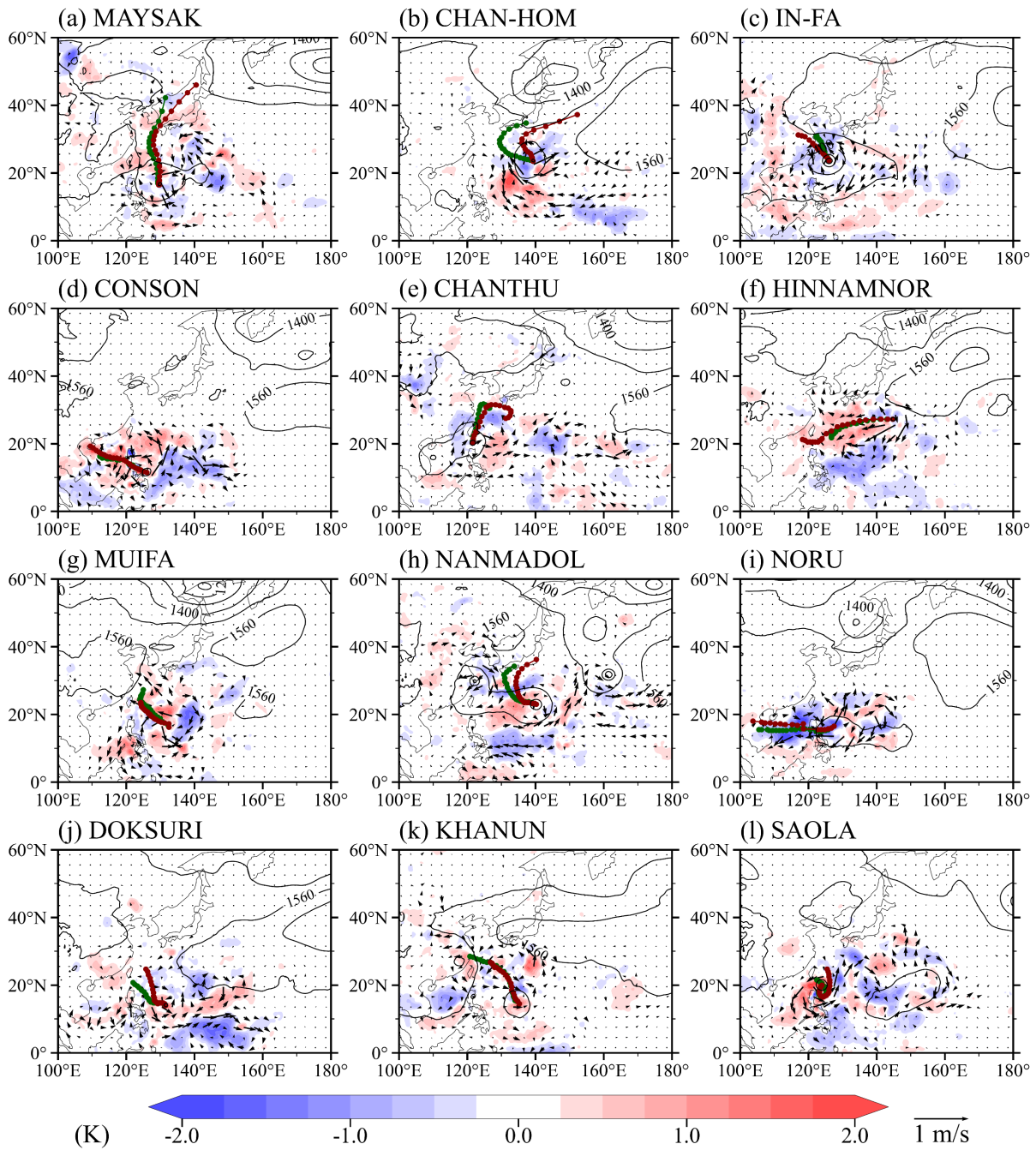

**Figure 10.** CNOPs of 12 TC cases in 850hPa temperature (shading) and wind (vector) perturbations. Contours represent geopotential heights. Green lines and red lines represent TC tracks from control forecast and perturbed forecast.

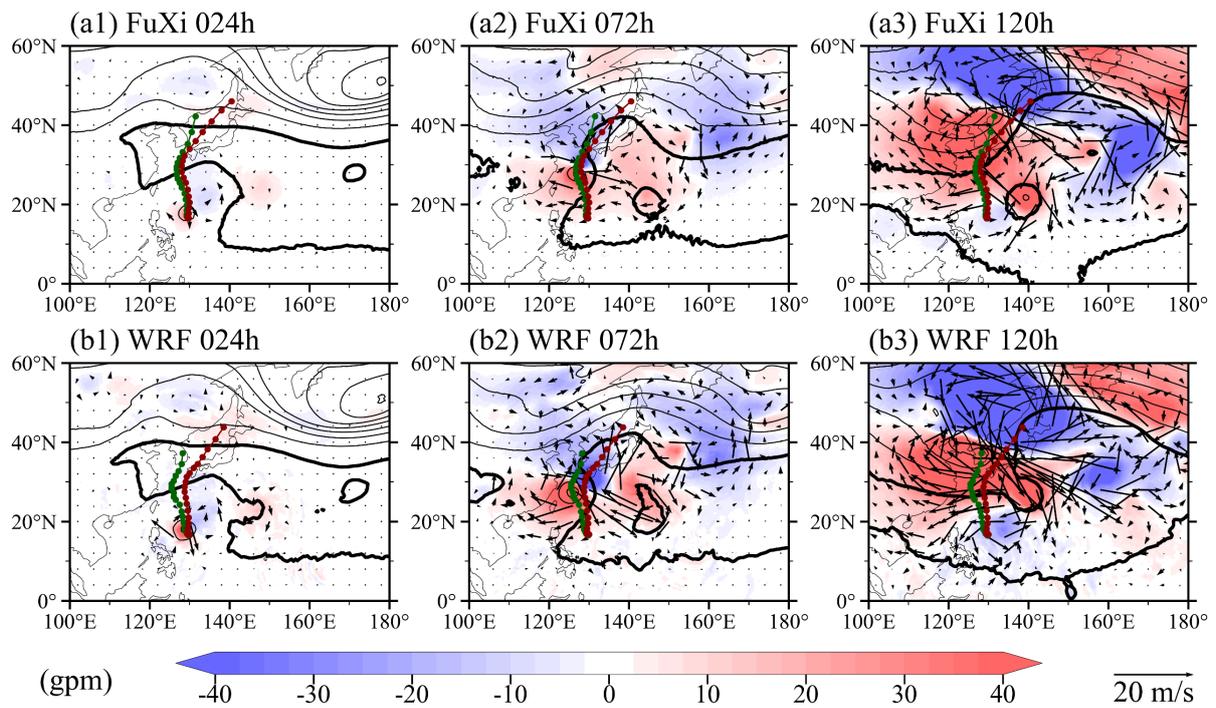

**Figure 11.** CNOP evolution of TC Maysak in FuXi and WRF. Figures (a1-a3) give the CNOP-caused 500hPa geopotential (shading) and wind (vectors) perturbations at 24h, 72h and 120h. Contours represent geopotential height from control forecast (interval at 80gpm, thickened at 5880gpm). Figures (b1-b3) are similar as Figures (a1-a3), but for results from WRF.

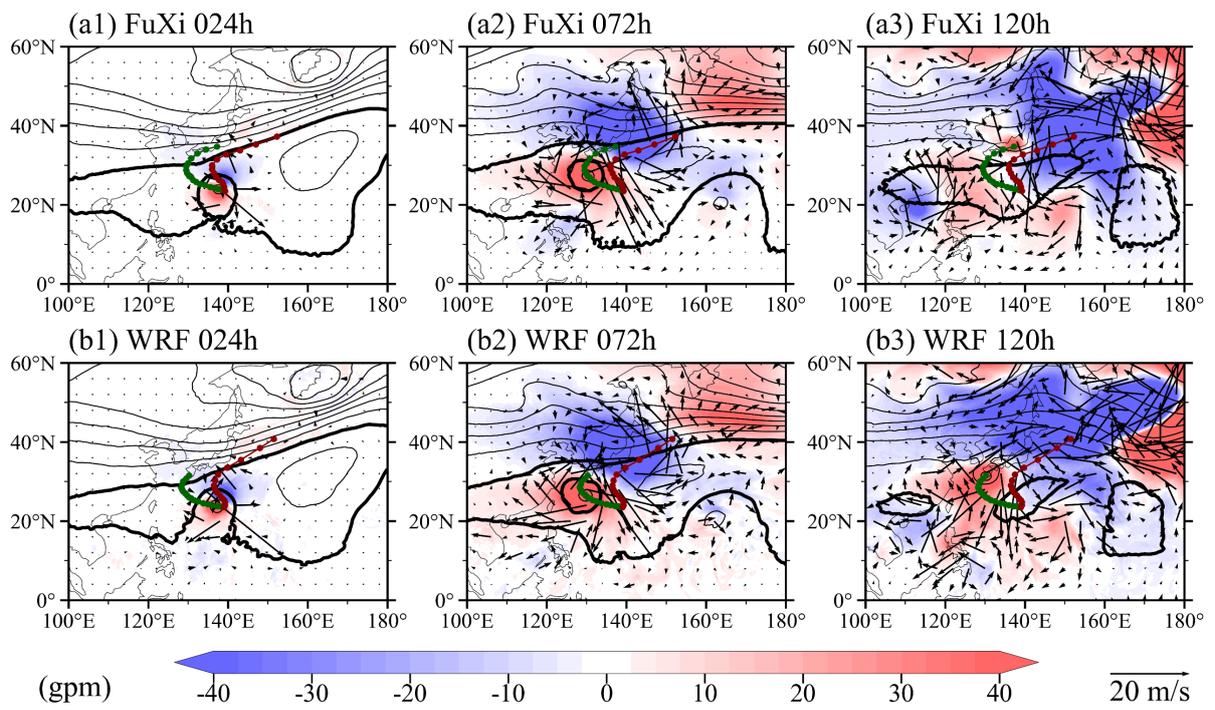

**Figure 12.** Same as Figure 11, but for TC Chanhom.

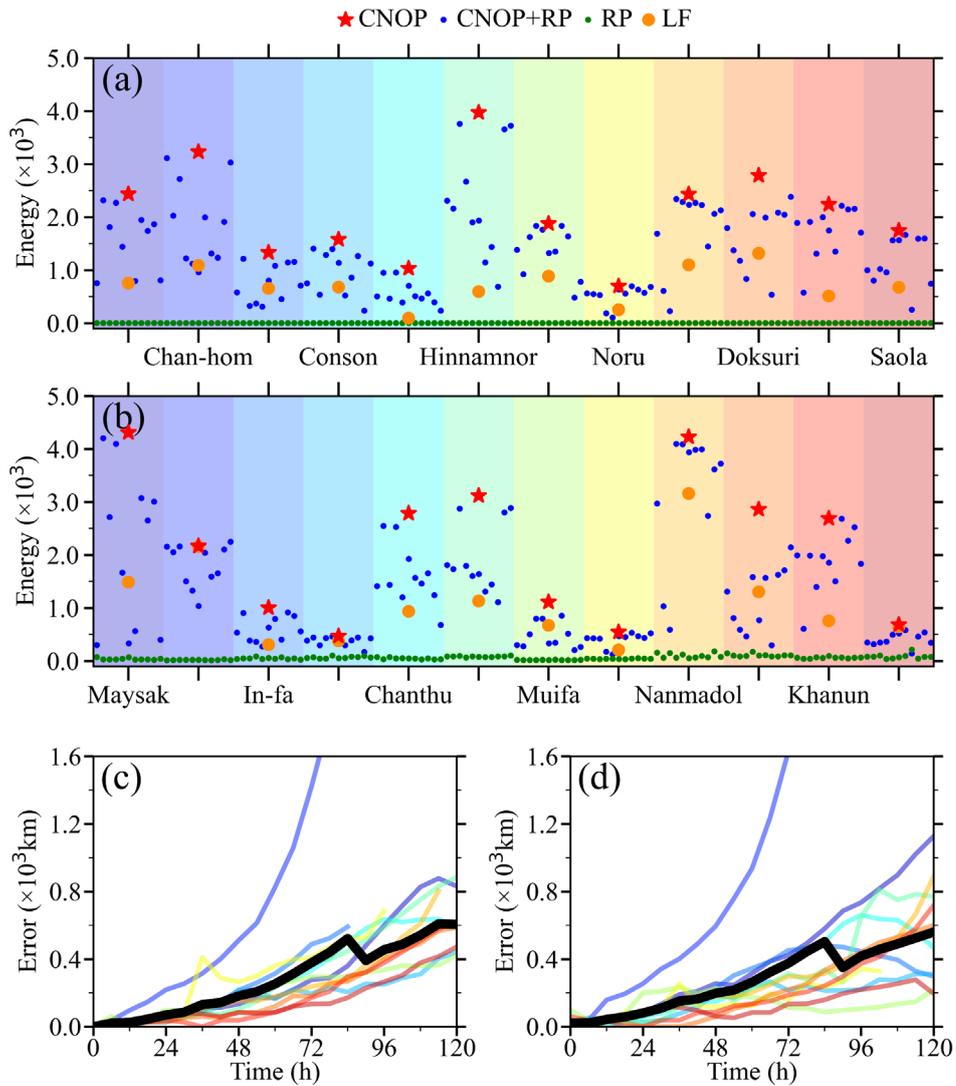

**Figure 13.** Target values and track error growth of different TC cases in FuXi and WRF. Figure (a) gives target values of CNOPs (red star), random perturbation (RP, green dots), lagged forecast perturbations (LFP, orange dots) and CNOPs combined with RPs (CNOP+RP, blue dots) evaluated from FuXi in different TC cases. Figure (b) is similar as Figure (a), but for results from WRF. Figure (c) gives CNOP-caused TC track errors of different cases (colored lines) and mean track errors (black line) in FuXi. Figure (d) is similar as Figure (c), but for results in WRF.

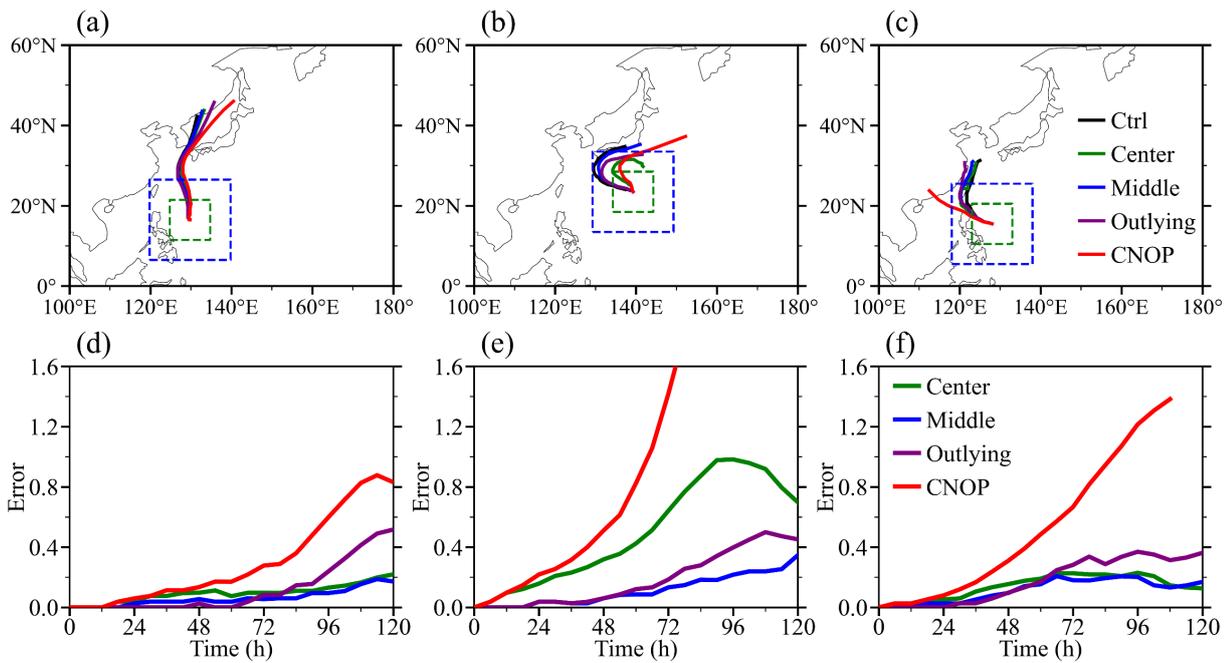

**Figure 14.** TC tracks from control and perturbed forecasts and their track errors. Figures (a, b, c) give the tracks of TC Maysak, Chanhom and Chanthu from control forecast (black) and forecasts perturbed by center (green), middle (blue), outlying (purple) and all (red) parts of CNOP. Figures (d, e, f) give the errors of those perturbed forecasts from control forecasts. Green (blue) squares in a-c show the boundary of the center parts (outlying parts).